\documentclass[a4paper, usenatbib]{mnras}

\usepackage[T1]{fontenc}
\usepackage{ae,aecompl}
\usepackage{graphicx}
\usepackage{mathrsfs}
\usepackage{amsmath}
\usepackage{amssymb}
\usepackage{verbatim}
\usepackage{multirow}
\usepackage{subfig}
\usepackage{mathtools}
\usepackage{multirow}
\usepackage[noabbrev]{cleveref}
\def\apjl{ApJ}
\def\apj{ApJ}                 
\def\apjl{ApJ}                
\def\apjs{ApJS}               
\def\ao{Appl.~Opt.}           
\def\aap{A\&A}                
\def\mnras{MNRAS}             
\def\pasj{PASJ}               

\def\memsai{Mem.~Soc.~Astron.~Italiana}

\def \inte {{\em INTEGRAL}}
\def \swift {{\em Swift}}

\def \xmm {{\em XMM-Newton}}
\def \rxte{{Rossi X-ray Timing Explorer}}

\def\swiftj{{SWIFT J1756.9$-$2508}}
\def\saxJ{{SAX J1748.9$-$2021}}
\def \saxj{SAX J1808.4$-$3658}
\def \igrj{IGR J00291+5934}

\def \swiftxrt{{\em Swift-XRT}}
\def \swiftbat{{\em Swift-BAT}}
\def \nustar{{\em NuSTAR}}
\def \rxte{{\em RXTE}}

\def \nicer{{\em NICER}}

\title[The 2018 outburst of \swiftj{}]{\swiftj{}: spectral and timing properties of its 2018 outburst}
\author[Sanna et al. ]{A. Sanna$^{1}$\thanks{E-mail:
    andrea.sanna@dsf.unica.it}, F. Pintore$^{2}$, A. Riggio$^{1}$, S.~M. Mazzola$^{3}$, E. Bozzo$^{4}$, T. Di Salvo$^{3}$ \newauthor 
    C. Ferrigno$^{4}$, A.~F. Gambino$^{3}$, A. Papitto$^{5}$, R. Iaria$^{3}$, L. Burderi$^{1}$\\
$^{1}$Dipartimento di Fisica, Universit\`a degli Studi di Cagliari, SP Monserrato-Sestu km 0.7, 09042 Monserrato, Italy\\
$^{2}$INAF-Istituto di Astrofisica Spaziale e Fisica Cosmica - Milano, via E. Bassini 15, I-20133 Milano, Italy\\
$^{3}$Universit\`a degli Studi di Palermo, Dipartimento di Fisica e Chimica, via Archirafi 36, 90123 Palermo, Italy\\
$^{4}$ISDC, Department of Astronomy, University of Geneva, Chemin d'\'Ecogia 16, CH-1290 Versoix, Switzerland\\
$^{5}$INAF-Osservatorio Astronomico di Roma, Via di Frascati 33, I-00078, Monte Porzio Catone (Roma), Italy}
\begin{document}

\date{Accepted 2018 August 18. Received 2018 August 18; in original form 2018 June 17}

\pagerange{\pageref{firstpage}$-$\pageref{lastpage}} \pubyear{}

\maketitle

\label{firstpage}

\begin{abstract}
We discuss the spectral and timing properties of the accreting millisecond X-ray pulsar \swiftj{} observed by \xmm{}, \nicer{} and \nustar{} during the X-ray outburst occurred in April 2018. The spectral properties of the source are consistent with a hard state dominated at high energies by a non-thermal power-law component with a cut-off at $\sim70$ keV. No evidence of iron emission lines or reflection humps has been found. From the coherent timing analysis of the pulse profiles, we derived an updated set of orbital ephemerides. Combining the parameters measured from the three outbursts shown by the source in the last $\sim11$ years, we investigated the secular evolution of the spin frequency and the orbital period. We estimated a neutron magnetic field of $3.1\times 10^{8}\,\,\, \textrm{G}<B_{PC}<4.5\times 10^{8}\,\,\, \textrm{G}$ and measured an orbital period derivative of $-4.1\times 10^{-12}$ s s$^{-1}$ $<\dot{P}_{orb}<7.1\times 10^{-12}$ s s$^{-1}$. We also studied the energy dependence of the pulse profile by characterising the behaviour of the pulse fractional amplitude in the energy range 0.3--80 keV. These results are compared with those obtained from the previous outbursts of \swiftj{} and other previously known accreting millisecond X-ray pulsars. 
%
%

\end{abstract}
 

\section{Introduction}
\swiftj{} is a low-mass X-ray binary discovered on 2007 June 7 during an X-ray outburst observed by the \swiftbat{}. Follow-up observations carried out with the \swiftxrt{} and the {\it Rossi X-ray Timing Explorer} (\rxte{}) provided the localisation of the source with an arcsec accuracy and led to the discovery of pulsations at a frequency of $\sim182$~Hz, classifying the source as an accreting millisecond X-ray pulsar \citep[AMXP, see e.g.][for a review]{Patruno12b}, in a 54.7~minutes orbit \citep{Krimm07}. A second outburst was recorded in July 2009 and the result of the observational campaign carried out with \swift{} and \rxte{} was reported in \citet[][hereafter P10]{Patruno10b}. In both occasions, the source displayed a spectral energy distribution compatible with the so-called ``island/extreme island state'' of an atoll source \citep[see e.g.][and reference therein]{Hasinger1989aa} and reasonably well described by a model comprising a power-law with a photon index of $\Gamma$=1.8-2.0 with no high-energy cut-off and a black-body component with a temperature of $kT$=0.4-0.7 keV \citep{Linares08}. Based on the upper limits derived on the spin-down torque, the neutron star magnetic field was constrained in a range compatible with values  expected for an AMXP and observed from other sources of this class \citep[0.4$\times$10$^8$~G$<$B$<$9$\times$10$^8$~G;][]{Patruno10a}. 
The source was discovered to undergo a new outburst by INTEGRAL on 2018 April 1 \citep{mereminskiy18}. The event was confirmed by \swiftbat{}, and follow-up observations provided the detection of pulsations at the known spin period of the source and a preliminary description of its broad-band X-ray spectrum \citep{krimm18,bult18a,bult18b,cha18,mazzola18,kuiper18, Bult2018c}. 

In this work, we carried out spectral and coherent timing analysis of the 2018 outburst of \swiftj{}, using \inte{}, \xmm{}, \nustar{} and \nicer{} observations of the source. We updated the source ephemerides and investigated the orbital period evolution over a baseline of almost 11 years by combining the current results with those reported from previous outbursts. We also discuss the broad-band spectral properties of \swiftj{}.

\section[]{Observations and data reduction}
\subsection{XMM-Newton}
\label{sec:XMM}
\xmm{} observed \swiftj{} on 2018 April 8 (Obs.ID. 0830190401) for a total exposure time of $\sim$~66 ks. During the observation, the EPIC-pn (hereafter PN) camera was operated in {\sc timing} mode and {\sc burst} mode for $\sim$ 49 ks and $\sim$ 10 ks, respectively. The RGS instrument observed in spectroscopy mode during the entire observation, while the EPIC-MOS1 and EPIC-MOS2 were operated in {\sc full frame} and {\sc timing} mode, respectively. To perform spectral and timing analysis of the source we focused on the PN and MOS2 data (the limited statistics and time resolution of the MOS1 data did not provide a significant improvement in any of the results presented here and in the following sections). These were processed using the Science Analysis Software (SAS) v. 16.0.0 with the up-to-date calibration files and the RDPHA calibrations \citep[see e.g.][]{Pintore15a}. We filtered events within the energy range 0.3-10.0 keV, retaining single and double pixel events only (\textsc{pattern$\leq$4}). We extracted the source events for the PN and MOS2 using  RAWX=[29:45] and RAWX=[285:325], respectively. We filtered background events for the PN selecting RAWX=[3:5] and we checked that the selected background was not contaminated by the emission from the source. For the MOS2, we extracted the background using an empty circular region of radius 150'' from the MOS1 dataset. The mean PN and MOS2 observed count rates during the observation were $\sim22$ cts/s and $\sim4.5$ cts/s, characterised by a slow decreasing trend. The background mean count rate in the PN selected RAWX range is of the order of $\sim0.5$ cts/s (0.3-10.0 keV). Thermonuclear (Type-I) X-ray burst episodes \citep[see e.g.][for a review]{Strohmayer2010aa} were not detected in the EPIC data.
We extracted RGS data with standard procedures. We checked that the RGS1 and RGS2 spectra were consistent and then we merged them with the task {\sc rgscombine}.

\begin{figure}
\centering
\includegraphics[width=0.48\textwidth]{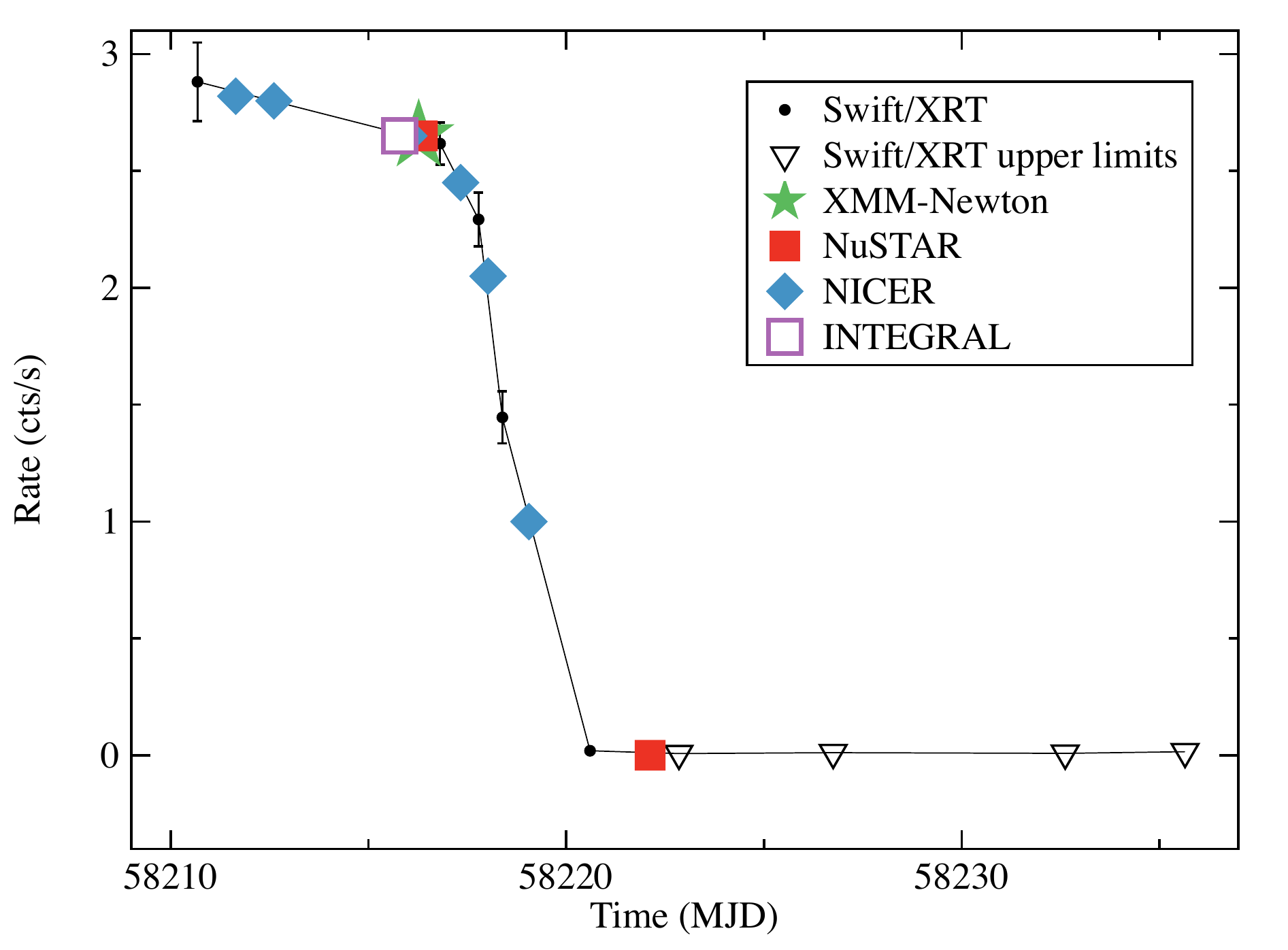}
\caption{\swiftxrt{} light curve (black points) of the 2018 outburst of the accreting millisecond X-ray pulsar \swiftj{}. Data are shown from MJD = 58210 (2018-04-02). Upper limits on the \swiftxrt{} count rate are shown with  empty triangles. The green star, red squares, blue diamonds, and purple square represent the starting times of the \xmm{}, \nustar{}, \nicer{} and \inte{} observations, respectively.}
\label{fig:lc}
\end{figure}

\noindent
Fig.~\ref{fig:lc} shows the monitoring light curve of the 2018 outburst of \swiftj{} as seen by \swiftxrt{} (black points) and obtained from the on-line \swiftxrt{} data products tool \citep{Evans2009a}. The green star represents the beginning of the \xmm{} observation taken few days after the outburst peak. To perform the timing analysis, we reported the PN photon arrival times to the Solar System barycentre by using the \textsc{barycen} tool (DE-405 solar system ephemeris). We applied the best available X-ray position of the source (reported in Tab.~\ref{tab:solution}) estimated performing astrometric analysis to the available \swiftxrt{} observation of the source \citep{Evans2009a}. The new source coordinates are compatible, to within the associated uncertainties, with the position reported by \citet{Krimm07}. 

\subsection{\nustar{}}
\nustar{} observed \swiftj{} twice during its 2018 outburst. The first observation (Obs.ID. 90402313002) started at 08:31 \texttt{UT} on 2018 April 8 for an elapsed time of $\sim 85$ks, resulting in a total effective exposure time $\sim43$ks. The second observation (Obs.ID. 90402313004) started at 02:56 \texttt{UT} on 2018 April 14 for an elapsed time of $\sim 125$ks, corresponding to a total effective exposure time of $\sim68$ks. 
The epochs at which \nustar{} observed are shown as red squares in Fig.~\ref{fig:lc}. We screened and filtered the events with the \nustar{} data analysis software (\textsc{nustardas}) version 1.5.1. We extracted the source events from the FPMA and FPMB focal planes within a circular region of radius 90$''$ centered on the source position. A  similarly extended region shifted to a position not contaminated by the source emission was used for the extraction of the background events. For each of the two observations, we obtained the background-subtracted light curves. These are characterised by an average count rate per FPM of $\sim10$ and $\sim 0.001$ counts/s, respectively. During the second observation the source was not significantly detected, and we thus discard these data for the remaining analysis.  We corrected the photon arrival times for the motion of the Earth-spacecraft system with respect to the Solar System barycentre with the {\sc barycorr} tools (using DE-405 solar system ephemeris), in analogy to what was done for the \xmm{} data.

\subsection{\nicer{}}

\nicer{} observed \swiftj{} seven times during its 2018 outburst (see Tab.~\ref{tab:obs} for more details). We extracted events across the 0.2-12 keV band applying standard screening criteria using the HEASOFT version 6.24 and NICERDAS version 4.0. Observations 105023105/6/7 showed the presence of high-energy background features. To further proceed with the timing analysis we excluded (when available) data 50 s before the raise and 100 s after the decay of the flares. We then barycentered the NICER photon arrival times with the {\sc barycorr} tool using DE-405 Solar system ephemeris and adopting the source coordinates reported in Tab.~\ref{tab:solution}.
\begin{table}
\begin{center}
    \begin{tabular}{ | c | c | c | c}
    \hline
    Instrument & Obs.ID. & Date & Exp. (s) \\ 
               & (revolution) &  & \\
    \hline
    \hline
    \xmm{}-PN & 0830190401 & 2018-04-08 & 49072 \\
    \hline
    \multirow{2}{*}{\nustar{}} & 90402313002 & 2018-04-08 & 43457 \\
     & 90402313004 & 2018-04-14 & 65763 \\
    \hline
    \inte{} & (1939) & 2018-04-07 & 85000 \\
    \hline
    \multirow{7}{*}{\nicer{}} & 1050230101 & 2018-04-03 & 6716 \\
     & 1050230102 & 2018-04-04 & 6424 \\
     & 1050230103 & 2018-04-07 & 2201 \\
     & 1050230104 & 2018-04-08 & 9490 \\
     & 1050230105 & 2018-04-09 & 3861 \\
     & 1050230106 & 2018-04-10 & 6141 \\
     & 1050230107 & 2018-04-11 & 4470 \\
    \hline
    \end{tabular}
    \caption{Log of the observations of \swiftj{} used to perform the spectral and timing analysis.}
    \label{tab:obs}
\end{center}
\end{table}
\subsection{INTEGRAL}
\label{sec:integral}

\swiftj{} was observed with \inte{} \citep{wink} from 2018 April 7 at 18:58 to 2018 April 8 at 19:56 (UTC), during the satellite revolution 1939. 
We analysed all data by using version 10.2 of the Off-line Scientific Analysis software (OSA) distributed 
by the ISDC \citep{courvoisier03}. 
The \inte{} observations are divided into science windows (SCWs), i.e. pointings with typical durations of $\sim$2-3~ks. 
We analysed a total of 25 SCWs in which the source was located to within an off-axis angle of 3.5~deg from the center of the JEM-X \citep{lund03} 
field  of view (FoV) and within an off-axis angle of 12~deg from the center of the 
IBIS/ISGRI \citep{ubertini03,lebrun03} FoV. These choices allowed us to minimise the instruments calibration 
uncertainties\footnote{http://www.isdc.unige.ch/integral/analysis}. 

We extracted first the IBIS/ISGRI and JEM-X mosaics.
\swiftj{} was detected in the IBIS/ISGRI 20-40~keV and 40-80~keV mosaics at a significance of 
$20\sigma$ and $13\sigma$, respectively. The corresponding fluxes estimated from the mosaics were 15.3$\pm$0.8~mCrab 
(roughly 1.2$\times$10$^{-10}$~erg~cm$^{-2}$s$^{-1}$) and 9.5$\pm$0.8~mCrab (roughly 7$\times$10$^{-11}$~erg~cm$^{-2}$s$^{-1}$).
The source was relatively faint for JEM-X and detected at $11\sigma$ in the 3-10~keV mosaic obtained by combining all JEM-X data. 
The correspondingly estimated flux was 26$\pm$3~mCrab (roughly 4.0$\times$10$^{-10}$~erg~cm$^{-2}$s$^{-1}$). We extracted the JEM-X light curves of the source with a bin time of 2~s to search for type-I X-ray bursts, but no significant detection was found.

\section{Data analysis and results}

\subsection{Spectral analysis}
We performed a broad-band spectral analysis combining all the available data. In particular, we selected the 2.0-10 keV, 1.2--2.0 keV, 3--70 keV, 30--90 keV and 4--40 keV for PN/MOS2, RGS, \nustar{} and \inte{}/ISGRI,  and JEMX, respectively.  

\begin{figure*}
\includegraphics[height=12cm,angle=270]{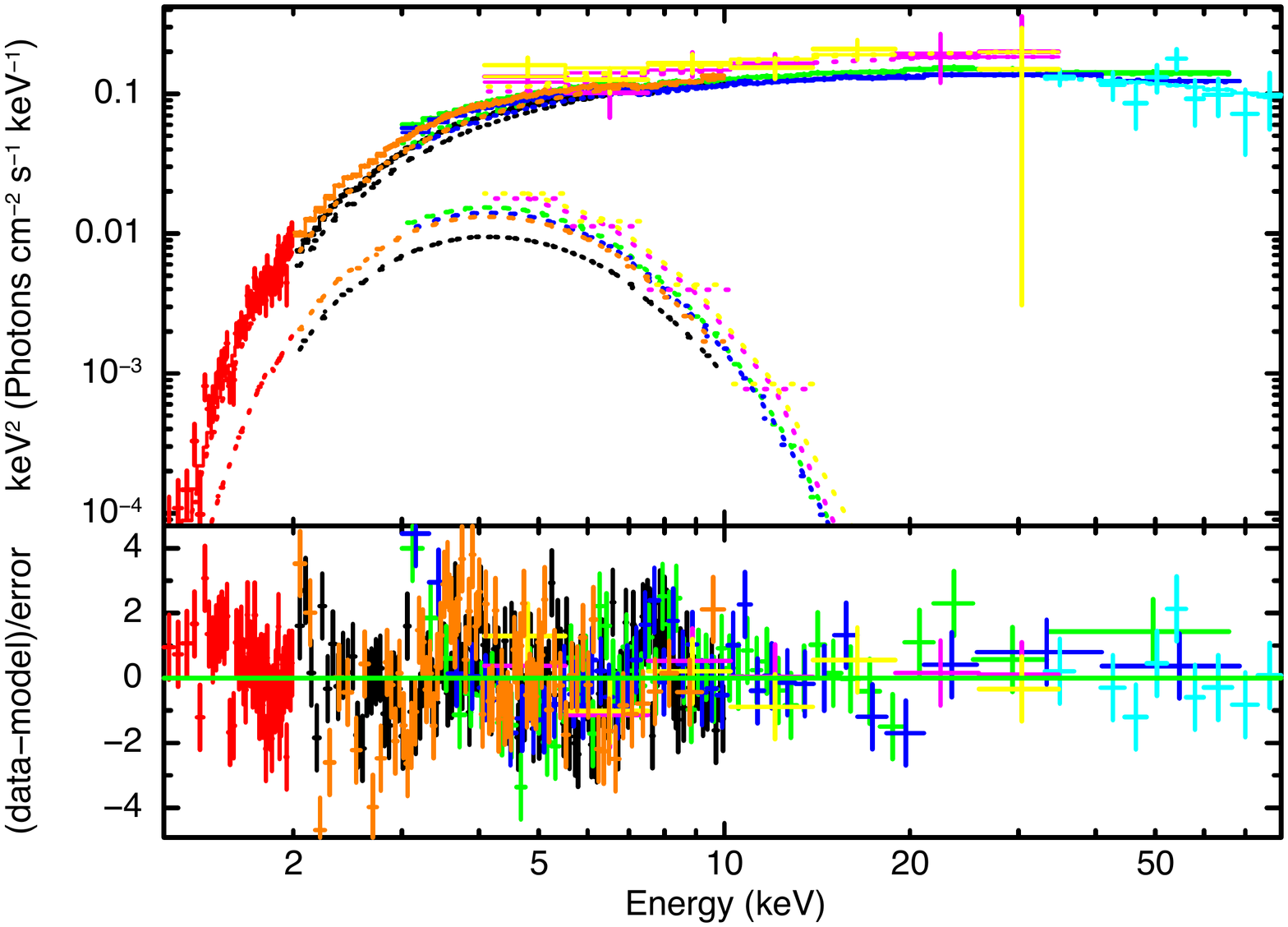}
\caption{\textit{Top panel:} Unfolded ($E^2f(E))$ X-ray spectrum of \swiftj{} observed by the PN (black), MOS2 (orange), RGS (red), \nustar{} FPMA/B (green and blue), JEMX (yellow) and ISGRI (cyan) fitted with the model {\sc tbabs$\times$(bbodyrad+cutoffpl)}, in the energy range 1--90 keV. \textit{Bottom panel:} residuals with respect to the best-fitting model. Data have been binned for displaying purposes only.}.
\label{spec11}
\end{figure*}

We first fitted  these spectra simultaneously adopting a simple {\sc tbabs*(cutoffpl)} model, with the addition of a multiplicative constant to take into account differences in the inter-calibrations of the instruments and the non-simultaneity between the datasets. 
The fit with this model did not provide an acceptable result ($\chi^2/dof=2903.49/1349$), showing a marked discrepancy between the spectral slopes of the PN and \nustar{} data. This is a well know issue and was already reported in the past \citep[see e.g.][]{Sanna2017b}. We thus allowed the photon indexes of the PN and \nustar{} spectra to vary independently in the fit. Although the fit was statistically improved  ($\chi^2/dof=1872.42/1348$), some residuals were still present and visible especially at the lower energies. We added a soft component (a multicolour black-body disc, {\sc diskbb} in {\sc xspec}; \citealt{Mitsuda84}) to the spectral model, which provided an additional significant improvement to the fit ($\chi^2/dof=1648.36/1346$). Assuming a distance of 8.5 kpc (based on the proximity toward the direction of  the Galactic center) and an inclination angle of $\leq60\deg$ inferred taking into account the lack of dips and eclipses in the X-ray light-curve \citep[see e.g.][]{Frank02}, we estimated an implausible inner disc radius of $\leq1.2$ km. We thus replaced the {\sc diskbb} component with a single-temperature {\sc bbodyrad}. The quality of the fit did not change significantly and we measured a black-body temperature of $0.85\pm0.03$ keV. The corresponding emitting radius was estimated at $1.8\pm0.2$ km, compatible with the size of an hot spot on the NS surface. The broad-band spectrum of \swiftj{} is shown in Figure~\ref{spec11}, together with the best fit model and the residuals from the fit.  All parameters of the best fit model are listed in Tab.~\ref{tab:spectrum}. We note that no emission lines were detected in the spectra, at odds with the findings reported from the analysis of X-ray data collected during the previous outbursts from the source (P10). The 3$\sigma$ upper limit that we obtained on the equivalent width of an iron emission feature centred at 6.5 keV and characterised by a width of 0.3 keV (see P10) for the 2018 outburst of the source is 5 eV.

\begin{table}
\begin{center}
\begin{tabular}{l | c  c  }
\hline
Model & Parameter & \\
\hline
{\sc tbabs} & nH ($10^{22}$) &  $8.14^{+0.14}_{-0.15}$\\
\multirow{ 2}{*}{{\sc bbodyrad}} & kT (keV) & $0.85^{+0.03}_{-0.04}$\\
& Norm & $4.5^{+0.8}_{-0.7}$ \\

\multirow{ 5}{*}{{\sc cutoffpl}}  & $\Gamma$(pn) & $1.44^{+0.04}_{-0.04}$ \\
 & $\Gamma$ (NuSTAR) & $1.65^{+0.03}_{-0.03}$\\
 & $\Gamma$ (MOS2) & $1.54^{+0.04}_{-0.04}$\\
 & E$_{cut}$ (keV) & $75^{+13}_{-10}$ \\
 & Norm & $0.042^{+0.003}_{-0.003}$ \\
\hline
& $\chi^2/dof$ & 1646/1345 \\
\hline
\end{tabular}
\caption{Spectral parameters obtained from the best fit to the \swiftj{} data of the 2018 outburst with the model described in the text ({\sc tbabs$\times$(bbodyrad+cutoffpl)}).}
\label{tab:spectrum}
\end{center}
\end{table}

\subsection{Timing analysis}
\label{sec:ta2018}
To investigate the timing properties of \swiftj{} we corrected the delays of the PN photon time of arrivals caused by the X-ray pulsar orbital motion under the hypothesis of a circular orbit. As a starting point, we considered the orbital period ($P_{\text{orb}}=3282.32(3)$ s and the projected semi-major axis (a$\sin(i)$/c=0.00598(2) lt/s) corresponding to the  ephemerides obtained from the 2009 outburst of the source (see Tab.~2 in P10). To investigate possible shifts on the time of passage from the ascending node ($T_{\text{NOD}}$), we extrapolated the closest value to the PN observation starting from the value reported in P10 and assuming a constant orbital period. We then explored a grid of parameters spaced by 1~s within a range of few kilo-seconds around the expected value. We searched for pulsations by exploiting the epoch-folding technique on the entire observations using 16 phase bins, starting with $\nu_0$ = 182.065803~Hz and exploring around $\nu_0$ with steps of $10^{-7}$ Hz, for a total of 10001 steps. The pulse profile with the largest signal-to-noise ratio was found at $\nu=182.065803(1)$~Hz and $T_{\text{NOD}}=58216.18423(1)$~MJD. 

Starting from the latter orbital solution, we corrected the photon arrival times in the PN and \nicer{} observations and we created pulse profiles by epoch-folding 500~s-long data segments using 8 phase bins.  As a starting point, we used the mean spin frequency  $\nu=182.065803(1)$~Hz obtained from the preliminary analysis of the PN data. Close to the tail of the outburst, we increased the length of the data segments in order to obtain statistically significant pulse profiles. Each pulse profile was modelled with a sinusoid from which we measured the amplitude and the fractional part of the phase residual. We retained only profiles with ratio between the sinusoidal amplitude and the corresponding $1\sigma$ uncertainty equal or grater than 3. The addition of a second harmonic did not improve the fit to the pulse profiles, being statistically significant in less than 20\% of the  intervals.

To improve the source ephemeris, we carried out a coherent timing analysis on the combined PN and \nicer{} data by fitting the time evolution of the pulse phase delays with the model: 
\begin{eqnarray}
\label{eq:ph}
\label{eq:ph_fit}
\Delta \phi(t)=\phi_0+\Delta \nu\,(t-T_0)+\frac{1}{2}\dot{\nu}\,(t-T_0)^2+R_{orb}(t),
\end{eqnarray}
where $\phi_0$ represents a constant phase, $\Delta \nu$ is a correction factor on the frequency used to epoch-fold the data, $\dot{\nu}$ represents the spin frequency derivative determined with respect to the reference epoch ($T_0$), and $R_{orb}(t)$ is the residual orbital modulation caused by discrepancies between the \emph{real} set of orbital parameters and those used to correct the photon time of arrivals \citep[see e.g.,][]{Deeter81}.

\begin{figure}
\centering
\includegraphics[width=0.48\textwidth]{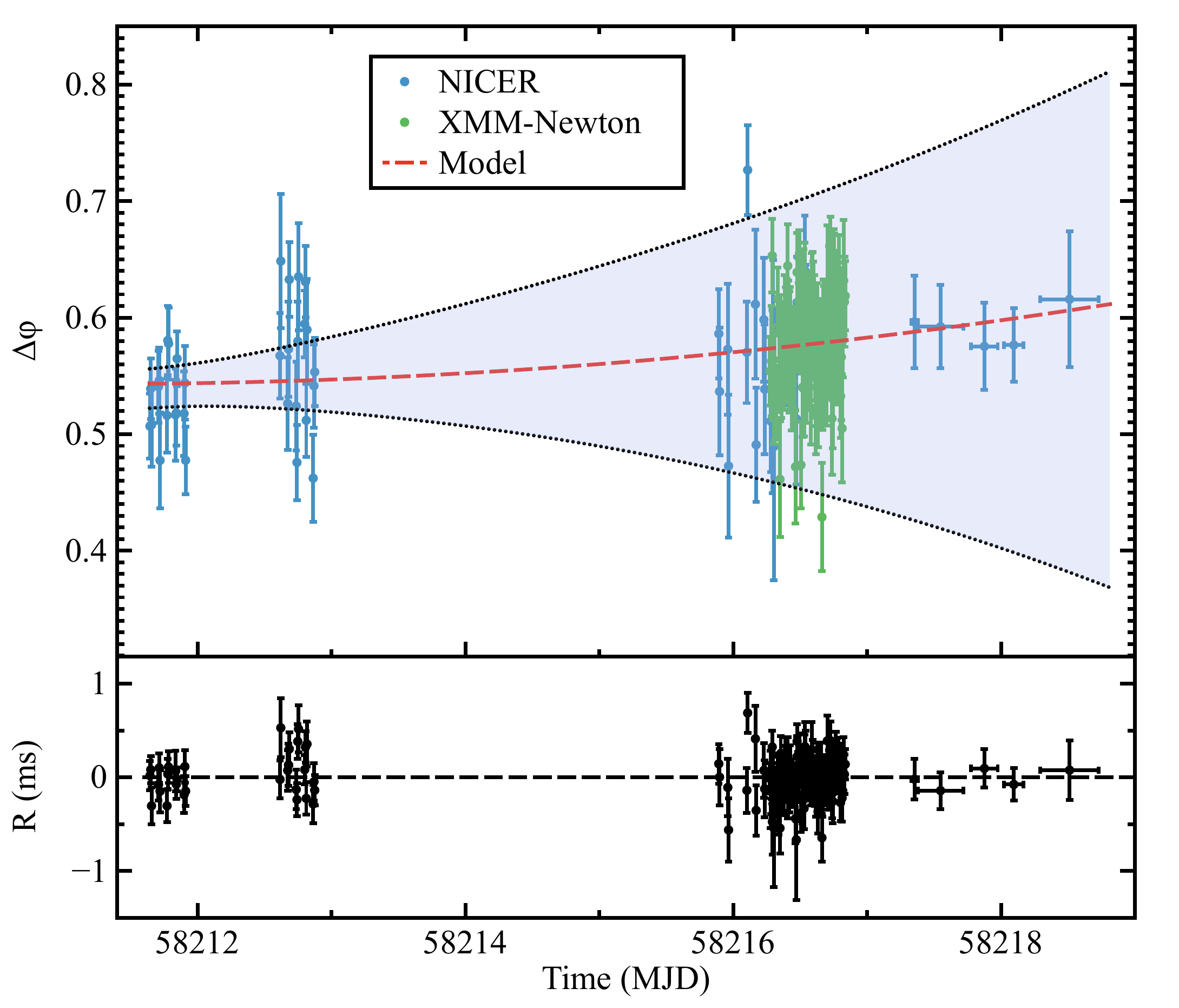}
\caption{\textit{Top panel:} Evolution of the pulse phase delays obtained by epoch-folding 500~s-long intervals of PN and \nicer{} data (shown in blue and green, respectively). Data are shown from MJD $\simeq$ 58211.6 (2018-04-03 14:24:00.0 UTC). The red dotted line represents the best-fit model described in the text, while the light-blue shaded area delimited by the black dotted lines represents the 95\% confident region. \textit{Bottom panel:} Residuals in ms with respect to the best-fitting model for the pulse phase delays.}
\label{fig:phase_fit}
\end{figure}

For each new set of orbital parameters obtained from this analysis, we applied the corrections to the photon arrival times and created new pulse phase delays that we modelled with Eq.~\ref{eq:ph}. We iteratively repeated this process until no significant improvements were found for any of the model parameters. We reported the best-fit parameters in the left column of Tab.~\ref{tab:solution}, while in Fig.~\ref{fig:phase_fit} we show the pulse phase delays for the PN and \nicer{} with the best-fitting models.  We note that the aforementioned timing solution is compatible within the errors with that reported by \citet{Bult2018c} from the analysis of the \nicer{} dataset only.

Using the updated set of ephemerides reported in Tab.~\ref{tab:solution}, we corrected the times of the \nustar{} events and we epoch-folded 800~s-long intervals. We modelled the pulse profiles with a sinusoidal model and we investigated the evolution of the pulse phase delays using Eq.~\ref{eq:ph}. The best-fit parameters, compatible within the uncertainties with those obtained from the phase-connected timing analysis of the PN and \nicer{} observations, are shown in the right column of Tab.~\ref{tab:solution}.  
\begin{table*}

\begin{tabular}{l | c  c  }
Parameters             & PN-\nicer{} & \nustar{}  \\
\hline
\hline
R.A. (J2000) &  \multicolumn{2}{c}{$17^h56^m57^s.43$}\\
DEC (J2000) & \multicolumn{2}{c}{$-25^\circ06'27''.4$}\\
Orbital period $P_{orb}$ (s) & 3282.40(4) & 3282.4(6)\\
Projected semi-major axis a sin\textit{i/c} (lt-ms) & 5.96(2)& 5.98(5) \\
Ascending node passage $T_{\text{NOD}}$ (MJD) & 58216.18433(10) & 58216.1841(2)\\
Eccentricity (e) & < 2$\times 10^{-2}$ & < 5$\times 10^{-2}$\\
\hline
\hline
Spin frequency $\nu_0$ (Hz) &182.06580377(11)& 182.065803(1)\\
Spin frequency 1st derivative $\dot{\nu}_0$ (Hz/s) &$<|1.4|\times 10^{-12}$ & $-4.3(2.1)\times 10^{-11}$\\
$\chi^2$/d.o.f. & {131.2/126} & 109.1(65)\\
\hline
\end{tabular}
\caption{Orbital parameters of the AMXP \swiftj{} obtained by phase-connecting data from the PN and \nicer{} (left column) and \nustar{} (right column) obtained during the source outburst in 2018. The reference epoch for the solution is T$_0$=58211.6 MJD. Uncertainties are reported at 1$\sigma$ confidence level. The best determined position of the source in X-rays has an associated uncertainty of 0.5$''$ (see the text for more details).}
\label{tab:solution}
\end{table*}

\begin{figure}
\centering
\includegraphics[width=0.48\textwidth]{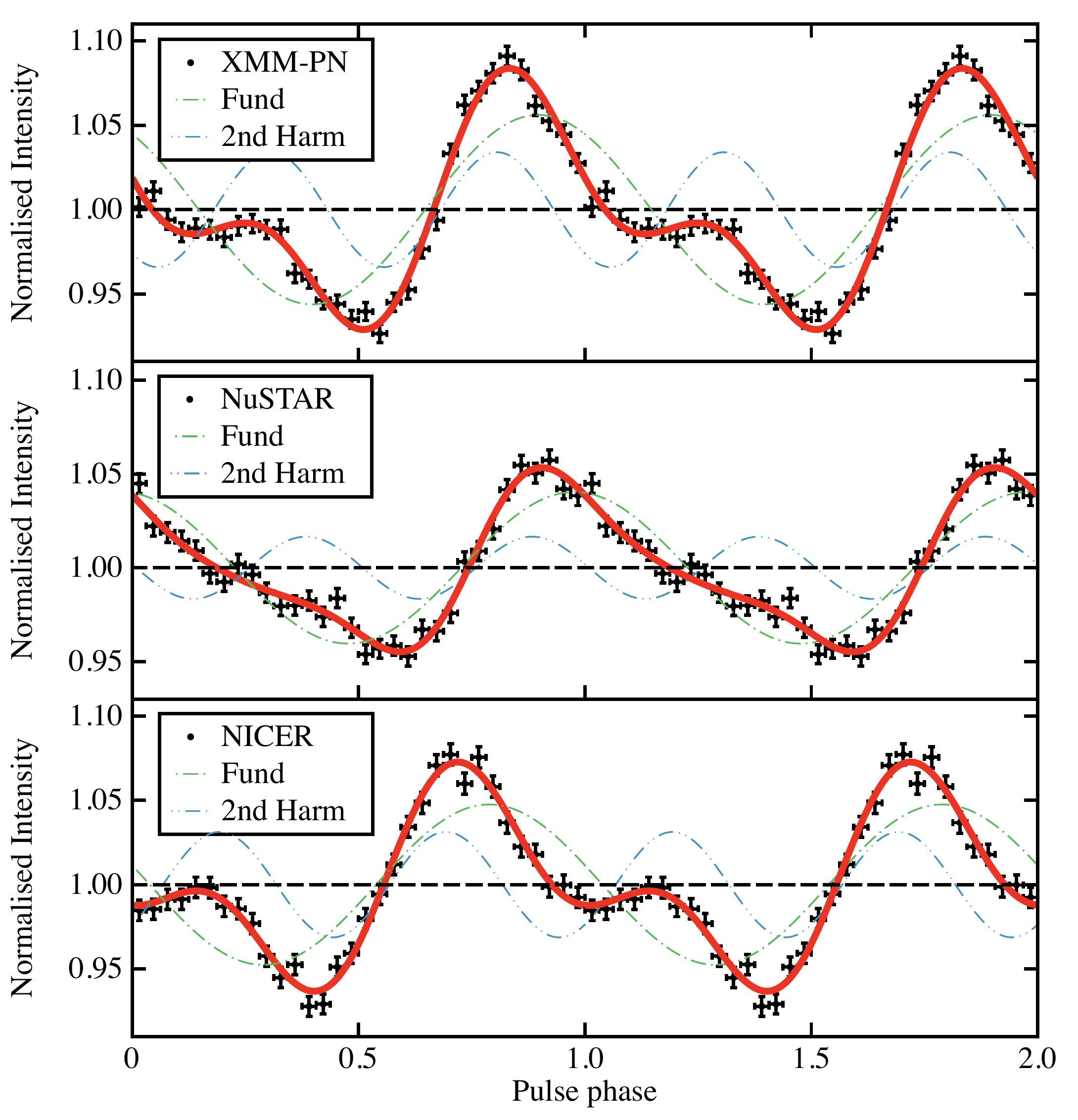}
\caption{\swiftj{} pulse profiles obtained by epoch-folding the \xmm{} (top panel), \nustar{} (medium panel), and \nicer{} (bottom panel) data. The updated set of ephemerides reported in Tab.~\ref{tab:solution} have been used together with a sampling of 32 bins. The best-fitting model, obtained by combining two sinusoids with harmonically related periods, is  reported in red. Green dot-dashed and blue dot-dot-dashed lines represent the fundamental and the second harmonic pulse profile components, respectively. We show in all cases two pulse cycles for clarity.}
\label{fig:pulse_prof}
\end{figure}
 
In Fig.\ref{fig:pulse_prof} we report the best pulse profiles obtained by epoch-folding the PN (top panel), \nustar{} (medium panel), and \nicer{} (bottom panel) data after correcting for the best-fitting parameters reported in Tab.~\ref{tab:solution}. The average pulse profile differs significantly from a sinusoidal function. It is well described by using a combination of two sinusoids shifted in phase. The \xmm{} (\nustar{}) fundamental and second harmonic have background-corrected fractional amplitudes of $\sim$5.6\% (4\%) and $\sim$3.4\% (1.6\%), respectively. For the \nicer{} average profile we obtain fractional amplitudes of $\sim$4.7\% and $\sim$3.1\% for the fundamental and second harmonic (not corrected for the background), respectively. 

\begin{figure}
\centering
\includegraphics[width=0.48\textwidth]{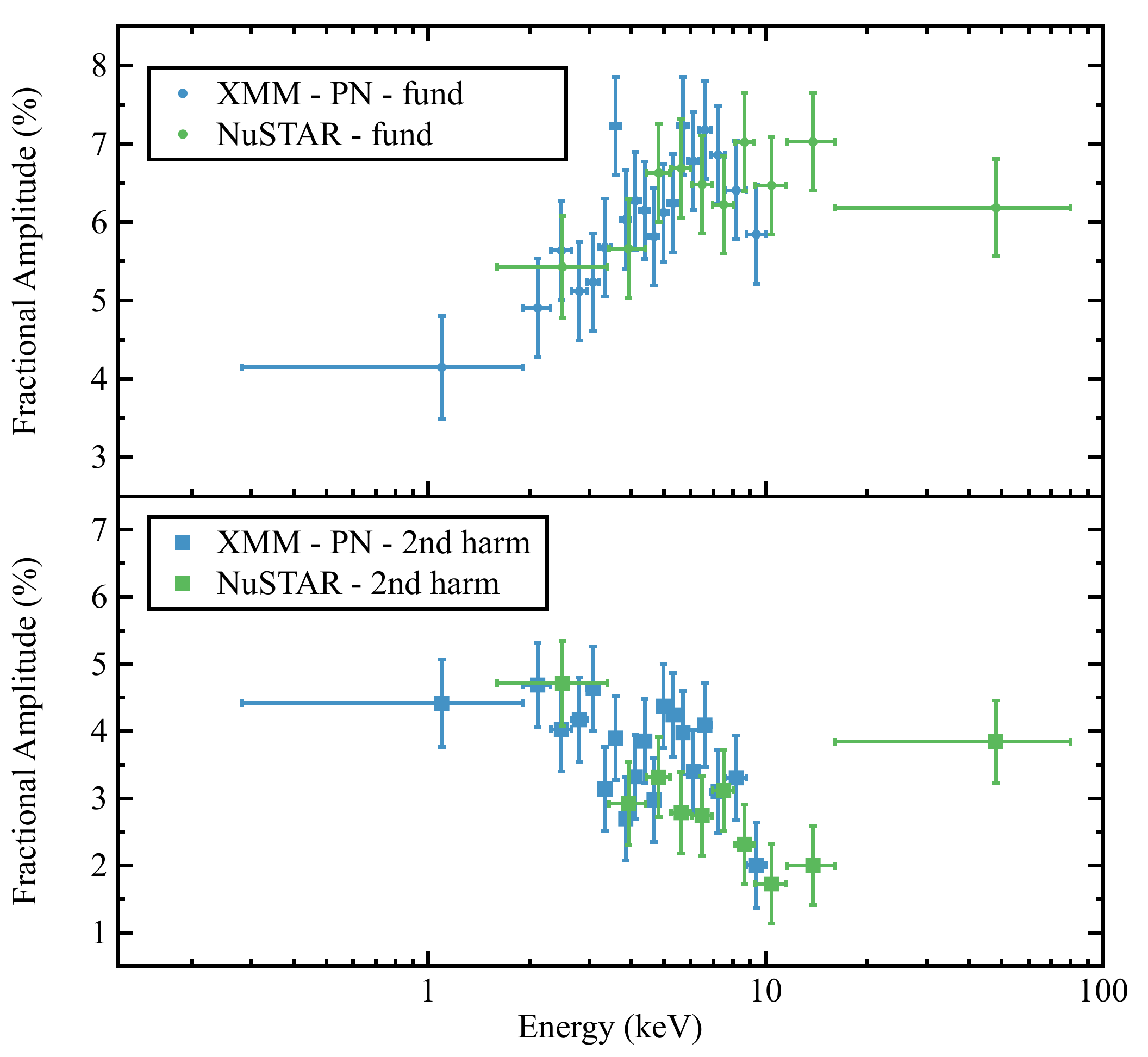}
\caption{Energy dependence of the pulse profile fractional amplitude for the fundamental (dots) and second harmonic (squares) components used to model the profiles obtained from the PN (blue) and the \nustar{} (green) datasets.}
\label{fig:amp_vs_energy}
\end{figure}

We also studied the energy dependence of the pulse profile by slicing the PN energy range (0.3--10 keV) in 20 intervals, and the \nustar{} energy range (1.6--80 keV) in 10 intervals. Energy bins have been selected in order to contain the same number of events. For each energy interval, we epoch-folded the events at the spin frequency values reported in Tab.~\ref{tab:solution} and we approximated the background-subtracted pulse profiles with a model consisting of two sinusoidal components (fundamental and second harmonic) for which we determined the fractional amplitudes. In Fig.~\ref{fig:amp_vs_energy}, we show the pulse profile energy dependence of the PN (blue) and \nustar{} (green) fractional amplitude for the fundamental (filled points) and second harmonic (filled squares) components. The PN fundamental component shows an increase from $\sim4\%$ at around 1 keV up to $\sim7\%$ at 6 keV, followed by a plateau around $\sim6\%$ above 10 keV. The second harmonic shows a decreasing trend (almost anti-correlated with the fundamental component) from $\sim4.5\%$ at 1 keV down to $\sim2\%$ at 10 keV. The \nustar{} fundamental component shows an increasing trend between $\sim5\%$ at around 2 keV and $\sim7\%$ up to 15 keV and then it stabilises up to 80 keV. Similarly to the PN data, the \nustar{} second harmonic decreases from $\sim5\%$ at 2 keV down to $\sim2\%$ at 10 keV where it starts increasing up to $\sim4\%$ at 80 keV. We note, however, that the statistics of the data is far from optimal and that future observations with an improved statistic combined with a finer sampling of the high energy region are needed to better investigate both the fundamental and second harmonic pulse fraction trend in this region.

\section{Discussion}

We presented the spectral and timing properties of the AMXP pulsar \swiftj{} obtained combining observations collected by \inte{}, \xmm{}, \nicer{} and \nustar{} during its 2018 outburst.

\subsection{Spectral properties}
The spectral results indicate that the source is highly absorbed ($N_{\rm H}$$\sim8.1\times10^{22}$ cm$^{-2}$), in line with previous findings \citep{Krimm07} and the \swiftxrt{} data taken at the beginning of the outburst \citep{mereminskiy18}. The $N_{\rm H}$ value we measured is slightly higher than that reported from the \nicer{} data \citep{bult18a}. Based on the comparison with other known AMXPs, we consider unlikely such large variability in the absorption column density and assume that the most reliable measure is provided by our spectral analysis (which includes data collected by the RGS instrument on-board \xmm{}).

The photon index of the power-law has changed from $2.04\pm0.03$ during the first days of the outburst (as measured from the \nicer{} data), to $\sim1.5$ during the \xmm{} observation. The detection of a cut-off at $\sim70$ keV strongly indicates that the source was in a hard state, as usually observed for AMXPs in outburst \citep[e.g.][ for a review]{Patruno12b, Burderi13}. Note that such a cut-off was not reported in the previous outburst of \swiftj{}, when the source spectrum displayed an hard tail extending up to 100 keV \citep{Linares08}.

At odds with previous outbursts (see P10), no significant iron lines were observed in the \nicer{} and \swiftxrt{} data collected during the event in 2018. We note, however, that the poor energy resolution of the \rxte{} data from the previous outbursts did not allow P10 to reliably constrain the line energy, the emissivity index, the inner disc radius, as well as the inclination of the system and the properties of the Compton reflection hump. This makes any comparison with the 2018 outburst particularly challenging. Assuming a line energy of 6.5 keV and a width of 0.3 keV (extrapolated from  the spectral residuals reported in Fig.~5 of P10), we estimated an upper limit on the equivalent width of any iron line not detected during the 2018 outburst of the order of 5 eV. We note that no evidence of iron emission lines or reflection humps has been reported also in the cases of the AMXPs IGR J16597$-$3704 \citep{Sanna2018a}, IGR J17379$-$3747 \citep{Sanna2018b}, XTE J1807$-$294 \citep[][]{Falanga05a}, and XTE J1751-305 \citep{Miller03}.
Finally, we measured an absorbed 0.3--70 keV flux of $(2.88\pm 0.01)\times 10^{-9}$ erg cm$^{-2}$ s$^{-1}$ (compatible with the flux values measured few days after the peak of 2007 and 2009 outbursts) and a luminosity of $2.5\times10^{36}$ erg s$^{-1}$, assuming a distance of 8.5 kpc (i.e., about 2$\%$ of the Eddington limit).

\subsection{Pulse profile and energy dependence}

We investigated the properties of the pulse profile of \swiftj{} as a function of energy by analysing the observations collected with the PN (0.3-10 keV) and \nustar{} (3-80 keV). Since the pulse profile is well described by a combination of two sinusoids (see Fig.~\ref{fig:pulse_prof}), we independently studied the fractional amplitude of the fundamental and second harmonic components. 

As reported in the top panel of Fig.~\ref{fig:amp_vs_energy}, the pulse fractional amplitude estimated from the fundamental component shows a clear increasing trend with energy, varying from 4\% to 7\% in the energy range 1-6 keV, followed by a slight decrease between 6 and 8 keV that at higher energies levels to $\sim 6$\%. This trend is very similar to that reported by P10 for the 2009 outburst, although we notice that the high energy behaviour of the fundamental component inferred from the \rxte{} data suggests a monotonic increase while the \nustar{} observation from the 2018 outburst clearly shows a constant tendency above 10 keV. Similar energy dependence of the fractional amplitude has been already reported for other AMXPs such as Aql X$-$1 \citep{Casella08}, \saxJ{} \citep{Patruno09a,Sanna2016a}, IGR J00291+5934 \citep[with a bit more complex energy dependence in the range 3-10 keV][]{Falanga05b,Sanna2017b} and XTE J1807$-$294 \citep{Kirsch04}. 
No consensus has been reached in terms of the process responsible for the hard spectrum of the pulsation detected in these sources. However, mechanisms such as a strong Comptonisation of the beamed radiation seems to well describe the properties of a few sources \citep{Falanga07b}. Nonetheless, it is worth noting that a completely opposite energy dependence of the pulse profile have been observed in other AMXPs such as XTE J1751$-$305 \citep{Falanga07b}, IGR J17511$-$3057 \citep{Papitto10,Falanga11,Riggio11a} and \saxj{} \citep{Cui98b,Falanga07b,Sanna:2017ab}.

Finally, the fractional amplitude of the second harmonic (Fig.~\ref{fig:amp_vs_energy} bottom panel), shows a clear decreasing trend from $\sim$4\% at 1 keV to $\sim$2\% at 10 keV, after which it starts increasing and reaches the values 4\% in the highest energy bin. We notice that the corresponding trend reported by P10 for the 2009 outburst shows a slightly weaker fractional amplitude below 10 keV.  

\subsection{Secular spin evolution}
\label{sec:spin}

\swiftj{} has been observed in outburst three times since its discovery \citep[see][]{Krimm07, Patruno10b}. To investigate the secular evolution of the spin frequency we considered the estimates from the 2007 and the 2009 \citep[reported by][see Tab~2 and 3]{Patruno10b}, that we combined with the 2018 spin frequency value reported in Tab.~\ref{tab:solution}.

\begin{figure}
\centering
\includegraphics[width=0.48\textwidth]{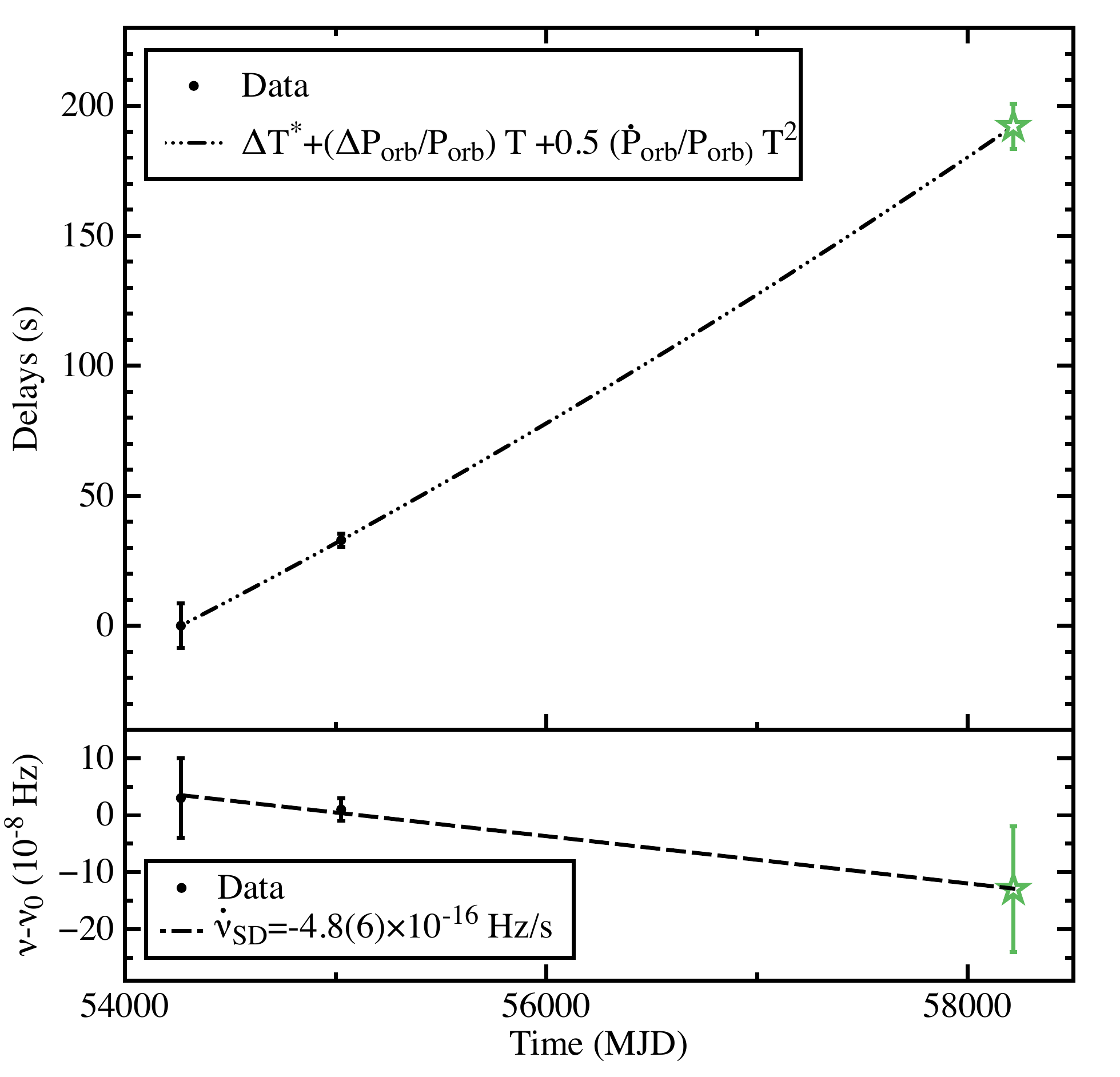}
\caption{{\it Top panel:}  Differential corrections to the time of passage at the ascending node for the three outbursts of \swiftj{}. Data are shown from MJD = 54000 (2006-09-22). The delays are calculated with respect to the first outburst of the source (see Sec.~\ref{sec:orb_ev} for more details). Black dots represent the delays from the first two outbursts reported by P10, while the green star is the value obtained combining the PN and \nicer{} observations. The dot-dot-dashed line represents the quadratic model that better describes the data. {\it Bottom panel:} Secular evolution of the spin frequency of \swiftj{} within the $\sim$11 yr baseline. Frequencies are rescaled with respect to the value $\nu_0 = 182.0658039$ Hz). Black points represent the frequency measurements of the previous outbursts estimated by P10, while the green star represents the spin values obtained from the combined timing analysis of the PN and \nicer{} observations. The dashed line represents the best-fitting linear model, corresponding to a spin down derivative $\dot{\nu}_{SD}=-4.8(6)\times 10^{-16}$~Hz/s, where uncertainties are reported at 1$\sigma$ confidence level.}
\label{fig:evo}
\end{figure}

We modelled the three spin frequency values with a linear function (see bottom panel in Fig.~\ref{fig:evo}), obtaining a best-fit with $\chi^2=0.03$ with 1 d.o.f. and a spin frequency derivative of $\dot{\nu}_{sd}=-4.8(6)\times 10^{-16}$ Hz/s, consistent with the upper limit reported by P10.   
Combining the rotational-energy loss rate to the rotating magnetic dipole emission, we can derive the magnetic field strength at the NS polar caps. Assuming a rotating dipole in presence of matter, the NS magnetic dipole moment can be approximated as
\begin{equation}
\mu \simeq 1.05\times10^{26}\left(\frac{1}{1+\sin^2{\alpha}}\right)^{-1/2} I_{45}^{1/2}\nu_{182}^{-3/2}\dot{\nu}_{-16}^{1/2}\,\,\,\, \text{G cm}^3,
\label{eq:mag}
\end{equation}
where $\alpha$ is the angle between the rotation and magnetic axes \citep[see e.g.][for more details]{Spitkovsky2006a}, $I_{45}$ is the NS moment of inertia in units of $10^{45}$ g cm$^2$, $\nu_{182}$ is the NS spin frequency rescaled for \swiftj{}, $\dot{\nu}_{-16}$ is the spin-down frequency derivative in units of $10^{-16}$ Hz/s. Adopting our estimates of the spin frequency and its secular spin-down derivative, and assuming the extreme values $\alpha=0$~deg and $\alpha=90$~deg we can limit the NS magnetic moment to be $2.3\times10^{26}\,\,\, \textrm{G\,cm$^3$}<\mu < 3.3\times10^{26}\,\,\, \textrm{G\,cm$^3$}$.
Defining the magnetic field strength at the magnetic caps as $B_{PC}= 2 \mu/R_{NS}^{3}$, and considering a NS radius of $R_{NS}=1.14\times10^{6}$ cm  \citep[corresponding to the FPS equation of state for a 1.4 M$_\odot{}$ NS, see e.g.,][]{Friedman1981a, Pandharipande1989a}, we obtain $3.1\times 10^{8}\,\,\, \textrm{G}<B_{PC}<4.5\times 10^{8}\,\,\, \textrm{G}$, consistent with the value reported by \citet{Mukherjee2015} and similar to what has been derived for other AMXPs. It is worth noting that the estimate presented here is likely a lower limit on the magnetic field strength. As a matter of fact, even though no significant spin-up derivative has been reported in the observed outbursts \citep[see e.g.][]{Krimm07, Patruno10b}, matter has been transferred and accreted on the NS surface likely accelerating the compact object.

\subsection{Orbital period evolution} 
\label{sec:orb_ev}

To investigate the secular evolution of the orbital period we  used the epoch of passage from the ascending node ($T_{\text{NOD}}$) measured in each of the three outbursts observed from the source, and the corresponding number of elapsed orbital cycles (N) determined with respect to a certain reference time at a specific orbital period. Under the assumption of constant orbital period, the predicted passages from the ascending node $T_{\text{NOD}_{{\text PRE}}}(N)=T_{\text{NOD}_{{\text 0}}}+N\, P_{orb}$ can be determined and compared with the measured ones  to calculate the corresponding differential corrections \citep[see e.g.][]{Papitto05, diSalvo08, Hartman08, Burderi09, Burderi2010a, Iaria2015a, Iaria2014a, Iaria2018a, Sanna2016a}. In order to be able to perform a coherent (orbital) timing, we need to verify that we can unambiguously determine the number of elapsed orbital cycles for each $T_{\text{NOD}}$. The condition is thus the following:
\begin{equation}
\left(\sigma^2_{T_{\text{NOD}}}+\sigma^2_{P_{orb}} N^2_{\text MAX}+\frac{1}{4}P^2_{orb} \dot{P}^2_{orb} N^4_{\text MAX}\right)^{1/2} < \frac{P_{orb}}{2}, 
\label{eq:orb_cycles}
\end{equation}
where $\sigma_{T_{\text{NOD}}}$ and $\sigma_{P_{orb}}$ are the uncertainties on the time of passage from the ascending node and the orbital period used as a reference for the timing solution, respectively. $\dot{P}_{orb}$ is the secular orbital derivative and $N_{\text MAX}$ is the integer number of orbital cycles elapsed by the source during the time interval covered by the three outbursts observed. Specifically, during the period 2007-2018, \swiftj{} elapsed $N_{\text MAX}=[(T_{\text{NOD}_{2018}}-T_{\text{NOD}_{2007}})/P_{orb}]\sim 104000$ orbital cycles.

Even considering the most accurate orbital period reported in Tab.~\ref{tab:par_fit_orb}, it is clear that despite the possible effects of an orbital period derivative, the condition reported in Eq.\ref{eq:orb_cycles} cannot be satisfied. It is thus not possible to unambiguously associate a number $N$ to all the $T_{\text{NOD}}$ values within the baseline 2007-2018. Instead, we can tentatively phase connect the first two outbursts separated in time by only 2.1 years, corresponding to $N_{\text MAX}\sim 2\times 10^4$. To test Eq.\ref{eq:orb_cycles}, we took the 2009 orbital period as a reference and we considered its uncertainty at the 95\% confidence level (0.06 s). Given the lack of knowledge on the orbital period derivative, we assumed as a fiducial value the average obtained combining the estimates from the only two AMXPs for which this quantity has been inferred: $\dot{P}_{orb}=3.6(4)\times 10^{-12}$  s/s for \saxj{} \citep[see e.g.][]{diSalvo08, Patruno2016a, Sanna:2017ab} and $\dot{P}_{orb}=1.1(3)\times 10^{-10}$  s/s for \saxJ{} \citep[][]{Sanna2016a}, that corresponds to $\dot{P}_{orb}\sim 6\times 10^{-11}$  s/s. Substituting the values into Eq.\ref{eq:orb_cycles}, we find the uncertainty on the time of passage from the ascending node to be of the order of $0.35P_{orb}$, which satisfies the possibility to apply coherent orbital timing on the two outbursts. Assuming $\dot{P}_{orb}\sim 6\times 10^{-11}$~s/s, we obtain an improved orbital period $P_{orb,2.1yr}=3282.3503(12)$~s. 

Substituting the more accurate estimate of the orbital period into Eq.\ref{eq:orb_cycles} and using the same prescription for $\dot{P}_{orb}$, we obtain that the propagated uncertainty on $T_{\text{NOD}}$ for the 2007-2018 baseline is below the $0.5P_{orb}$ threshold and we are then allowed to coherently phase connect the orbital parameters among the three outbursts.  
As we can unambiguously associate the number of elapsed orbital cycles to each $T_{\text{NOD}}$, we calculate the  correction on the NS passage from the ascending node $\Delta T_{\text{NOD}}$, with respect to the beginning of the 2007 outburst. For each outburst we determine the quantity $T_{\text{NOD}_{{\text obs}}}-T_{\text{NOD}_{{\text PRE}}}$ estimated with respect to $P_{orb,2.1yr}=3282.3503(12)$~s, and we plot it as a function of corresponding elapsed cycles (top panel in Fig.~\ref{fig:evo}). Using the quadratic function: 
\begin{eqnarray}
\label{eq:fit_tstar}
\Delta T_{\text{NOD}} = \delta T_{\text{NOD},2007} + N\, \delta P_{orb,2.1yr}+0.5\,N^2\, \dot{P}_{orb}P_{orb,2.1yr},
\end{eqnarray} 
we determine the unique set of parameters that approximate the $T_{\text{NOD}}$ values shown in Fig.~\ref{fig:evo}, where $\delta T_{\text{NOD} 2007}$ represents the correction to the adopted time of passage from the ascending node, $\delta P_{orb,2.1yr}$ is the correction to the orbital period, and $\dot{P}_{orb}$ is the orbital-period derivative. In the bottom part of Tab.~\ref{tab:par_fit_orb}, we report the combined orbital solution of the source and the corresponding uncertainties reported at the $1\sigma$ confidence level. 

\begin{table}
\begin{tabular}{cccc}
\hline
\hline
Outburst & $T_{\text{NOD}}$ & $P_{orb}$ & $\dot{P}_{orb}$ \\
 		 & (MJD) 			& (s) 		& ($10^{-12}$ s s$^{-1}$) \\
\hline
2007 & 54265.28087(10) & 3282.41(15) & - \\
2009 &  55026.03431(3) & 3282.32(3) & - \\
2018 &  58216.18433(10) & 3282.40(4) & - \\
\hline
Combined & 54265.28087(10) & 3282.3519(5) & $1.5 \pm 2.8$\\
\hline
\hline
\end{tabular}
\caption{{\it Top:} \swiftj{} best-fitting orbital parameters derived for each individual outburst. The values of the first two outbursts was obtained from P10. {\it Bottom:} Best-fitting orbital parameters derived combining the source outbursts observed between 2007 and 2018 (see text for more details). Uncertainties are reported at $1\sigma$ confidence level.}
\label{tab:par_fit_orb}
\end{table}

The uncertainty on the orbital period derivative is such that we cannot determine whether the orbit is secularly expanding or shrinking. However, the longer baseline with respect to P10 allow us to improve by few orders of magnitude the constraint on the strength of the orbital derivative. Already at this stage, we can exclude an orbital evolution similar to that of the AMXP \saxJ{} for which an extremely fast expansion has been reported \citep{Sanna2016a}. On the other hand, considering the 95\% confidence level interval $-4.1\times 10^{-12}$ s/s $<\dot{P}_{orb}<7.1\times 10^{-12}$ s/s \citep[see also][]{Bult2018c}, we note that the secular evolution of \swiftj{} is still compatible with the fast expansion reported for \saxj{} \citep[see e.g.][]{diSalvo08, Patruno2016a, Sanna:2017ab} as well as with very slow evolution suggested for \igrj{} \citep[see e.g.][]{Patruno2016b, Sanna2017b}. Future outbursts will allow us to further constrain the orbital period derivative and the secular evolution of the system.

\section{Conclusions}
We reported on the spectral and timing properties of the 2018 outburst of the AMXP \swiftj{} observed with \inte{}, \xmm{}, \nustar{} and \nicer{}. From the phase-connected timing analysis of the \nicer{} and \xmm{} observations, we obtained an updated set of the source ephemerides, compatible within the errors with those obtained from the \nustar{} dataset. Owing to the multiple observations performed during the source outburst, we obtained, for the first time since the decommission of \rxte{}, a reliable constraint on the spin frequency derivative (|$\dot{\nu}|<1.4\times 10^{-12}$ Hz/s) of an AMXP during the accretion state. Combing the timing properties from the previous two outbursts, we estimated a secular spin-down frequency derivative $\dot{\nu}_{sd}=4.8(6)\times 10^{-16}$ Hz/s, compatible with a magnetic field (at the polar caps) of $3.1\times 10^{8}\,\,\, \textrm{G}<B_{PC}<4.5\times 10^{8}\,\,\, \textrm{G}$. Furthermore, we obtained a secular orbital period derivative in the range $-4.1\times 10^{-12}$ s/s $<\dot{P}_{orb}<7.1\times 10^{-12}$ s/s (95\% confidence level), suggesting that more outbursts are required to further constrain the orbital evolution of the system. We also investigated the pulsation spectral energy distribution of \swiftj{} in the energy range 0.3--10 keV and 3--80 keV, using the \xmm{} and \nustar{} datasets, respectively. The pulse fractional amplitude trend shown by the fundamental and second harmonic components present similarities with those reported for other AMXPs likely suggesting a Comptonisation origin.  

Finally, we found that the broad-band (3--90 keV) energy spectrum of \swiftj{} observed during its 2018 outburst is well described by an absorbed cut-off power law plus a soft thermal component. A photon index of $\sim$1.5 combined with a cut-off at $\sim$ 70 keV strongly suggest that the source was observed in a hard state. Contrary to previous outbursts, we detected no significant reflection features, with a constraining upper limit on the iron line equivalent width ($\sim$ 5 eV).  

\section*{Acknowledgments}

We thank N. Schartel for providing us with the possibility to perform the ToO observation in the Director Discretionary Time, and the \xmm{} team for the technical support. We also use Director Discretionary Time on \nustar{}, for which we thank Fiona Harrison for approving and the \nustar{} team for the technical support. We acknowledge financial contribution from the agreements ASI-INAF I/037/12/0 and ASI-INAF 2017-14-H.O. This work was partially supported by the Regione Autonoma della Sardegna through POR-FSE Sardegna 2007-2013, L.R. 7/2007, Progetti di Ricerca di Base e Orientata, Project N. CRP-60529. AP acknowledges funding from the European Union's Horizon 2020 research and innovation programme under the Marie Sk\l{}odowska-Curie grant agreement 660657-TMSP-H2020-MSCA-IF-2014, as well as the International Space Science Institute (ISSIBern) which funded and hosted the international team ``The disk magnetosphere interaction around transitional millisecond pulsar''.

\bibliographystyle{mnras}
\bibliography{biblio}

\label{lastpage}

\end{document}